\documentclass[12pt,onecolumn,manuscript, tighten,trackchanges]{aastex631}
\usepackage{natbib}
\usepackage{multirow,color}
\usepackage{bbding}
\usepackage{amssymb}
\usepackage{ulem}
\usepackage{color}
\usepackage{longtable}
\usepackage{array}
\usepackage{bm}
\usepackage{url}
\usepackage{hyperref}
\usepackage{color}
\usepackage{graphicx}

\def\gsim{\;\lower4pt\hbox{${\buildrel\displaystyle >\over\sim}$}\;}
\def\lsim{\;\lower4pt\hbox{${\buildrel\displaystyle <\over\sim}$}\;}
\def\grls{\;\lower4pt\hbox{${\buildrel\displaystyle >\over <}$}\;}

\bibpunct{(}{)}{,}{o}{}{,} 

\begin{document}

\title{Why could a new-born active region produce coronal mass ejections?}  

\author{Hanzhao Yang}
\affiliation{Planetary Environmental and Astrobiological Research Laboratory (PEARL), School of Atmospheric Sciences, Sun Yat-sen University, Zhuhai, Guangdong, 519082, People's Republic of China}

\author{Lijuan Liu}
\affiliation{Planetary Environmental and Astrobiological Research Laboratory (PEARL), School of Atmospheric Sciences, Sun Yat-sen University, Zhuhai, Guangdong, 519082, People's Republic of China}
\affiliation{CAS center for Excellence in Comparative Planetology, People's Republic of China}
\affiliation{Key Laboratory of Tropical Atmosphere-Ocean System, Sun Yat-sen University, Ministry of Education, Zhuhai, People's Republic of China}
{\correspondingauthor{Lijuan Liu}}
{\email{liulj8@mail.sysu.edu.cn}}

\begin{abstract}

Solar active regions (ARs) are the main sources of flares and coronal mass ejections (CMEs). 
NOAA AR 12089, which emerged on 2014 June 10, produced two C-class flares accompanied by CMEs within five hours after its emergence.
When producing the two eruptive flares, the total unsigned magnetic flux ($\Phi_{\mathrm{AR}}$) and magnetic free energy ($E_f$) of the AR are much smaller than the common CME-producing ARs. 
Why can this extremely small AR produce eruptive flares so early? 
We compare the AR magnetic environment for the eruptive flares to that for the largest confined flare from the AR.
Besides the $\Phi_{\mathrm{AR}}$ and $E_f$, we calculate the ratio between the mean characteristic twist parameter ($\alpha_{\mathrm{FPIL}}$) within the flaring polarity inversion line (FPIL) region and $\Phi_{\mathrm{AR}}$, a parameter considering both background magnetic field constraint and non-potentiality of the core region, for the three flares. 
We find higher $\alpha_{\mathrm{FPIL}}/{\Phi_{\mathrm{AR}}}$ values during the eruptive flares than during the confined flare. 
Furthermore, we compute the decay index along the polarity inversion line, revealing values of 1.69, 3.45, and 0.98 before the two eruptive and the confined flares, respectively. 
Finally, nonlinear force-free field extrapolation indicates that a flux rope was repeatedly formed along the FPIL before eruptive flares, which ejected out and produced CMEs.
No flux rope was found before the confined flare. 
Our research suggests that even a newly emerged, extremely small AR can produce eruptive flares if it has sufficiently weak background field constraint and strong non-potentiality in the core region.

\end{abstract}

\section{Introduction}\label{sec:intro}

Solar flares and coronal mass ejections (CMEs) are the most energetic phenomena in the solar system, which can cause hazardous space weather near the geospace~\citep{Gosling_1993}.
The origin of flares and CMEs has been a hotspot of space and solar physics research for decades, yet it remains not fully understood.
The standard flare model is developed to describe the overall evolution of flares and CMEs~\citep[see review in][]{Shibata_etal_1995, Hudson_2001}.
In this model, flares and CMEs are thought to be different manifestations of the same process, i.e., the eruption of a magnetic flux rope.
However, observations suggest that solar flares are not always accompanied by CMEs.
Defining flares without CMEs as ``confined flares'' and flares with CMEs as ``eruptive flares'', it is found that confined flares are common in weaker events.
At the same time, for large flares, the association rate between flares and CMEs rises noticeably~\citep{Andrews_2003, Yashiro_2005, Yashiro_2006}.

Active regions (ARs) are the main sources of solar flares and CMEs. 
However, the flare and CME productivity of the ARs varies greatly.
A large amount of progress has been made in the past few decades in understanding what factors can determine the productivity of ARs.
It is suggested that the non-potentiality of the AR, e.g., magnetic helicity, magnetic twist, and magnetic free energy is correlated with its flare and CME
productivity~\citep[e.g.,][]{Nindos_2004, Guo_etal_2010, Vasantharaju_etal_2018, Thalmann_etal_2019, Vemareddy_2019, Gupta_etal_2021, Liu_etal_2023}.
Moreover, the eruptive characteristic of the flares may be governed by the constraining effect of the background magnetic field, e.g., the strength of the overlying background magnetic field and its decay with height~\citep[e.g.,][]{Kliem_2006, Wang_2007, Cheng_etal_2011, Filippov_2020, Mitra_etal_2022}.
The decline of overlying magnetic field strength with height can be quantified with a parameter called decay index.
This parameter is relevant to the torus instability, an ideal magnetohydrodynamic (MHD) instability~\citep{Kliem_2006}.
According to the theory, when the decay index reaches a critical value, the torus instability of a magnetic flux rope will set on, resulting in an eruption~\citep{Kliem_2006, Fan_2007, Demoulin_2010}.
Both magnetic non-potentiality and background field constraint may work together to influence the productivity of ARs~\citep{Nindos_2004, Liu_etal_2016, Vasantharaju_etal_2018}.

Compared to the parameters that quantify a single aspect of properties of the AR, \citet{Sun_2015} suggests using a relative measure that quantifies the ratio of magnetic non-potentiality to the constraint of background field to measure the AR productivity.
They have designed a ``flaring polarity inversion line'' (FPIL) mask to demarcate the AR core field where the most magnetic free energy resides.
Following the idea of~\citet{Sun_2015},~\citet{Li_etal_2022} found a new parameter $\alpha_{\mathrm{FPIL}}/{\Phi_{\mathrm{AR}}}$, i.e., the mean characteristic twist parameter ($\alpha_{\mathrm{FPIL}}$) within the FPIL mask region versus the total unsigned magnetic flux of the AR ($\Phi_{\mathrm{AR}}$), is well able to distinguish the capability of an AR to produce eruptive flares.

Despite the specific factors, generally, an AR with a larger area, stronger magnetic field, and more complex morphology is more likely to be CME-rich~\citep[e.g.,][]{Chen_etal_2011}.
Those ARs that can produce eruptive flares usually have a total unsigned magnetic flux exceeding the order of magnitude of $10^{22}$ Mx, and a magnetic free energy larger than the order of $10^{31}$ erg~\citep{Chen_etal_2011, Vasantharaju_etal_2018}.
However, we found an unusual AR here, NOAA AR 12089, which emerged on 2014 June 10 and produced two CMEs associated with C-class flares within five hours after the onset of its emergence. 
It produced 11 C-class and 1 M-class flares during the entire transit.
Except the first two, all subsequent flares were confined.
The AR is extremely small when producing the two eruptive flares, with both the total unsigned magnetic flux and magnetic free energy being an order of magnitude smaller than those of common CME-producing ARs~\citep{Chen_etal_2011, Vasantharaju_etal_2018}.
Moreover, the AR produced CMEs within 5 hours after its emergence, while a survey about 19 emerging ARs from their birth until they produced their first major flares or CMEs shows that of the 15 ARs that produced CMEs, all except AR 12089 produced CMEs about at least 18 hours after the onset of their emergence~\citep{Liu_etal_2021}.
As a rarely observed case, the AR provides a good opportunity for studying the generation of CMEs in a new-born, extremely small AR.
We compare the AR magnetic environment during the two C-class eruptive flares to that of the largest confined flare (M1.1 class) here. 
In Section~\ref{sec:data}, we describe the data and methods we used for investigation. 
In Section~\ref{sec:result}, we present the details of flares, the magnetic parameters about the AR productivity, and the evolution of flux ropes. 
In Section~\ref{sec:sum}, we provide a summary and discussion.

\section{DATA AND METHODS}\label{sec:data}

We examine NOAA {\it GOES} soft X-ray (SXR) reports to search for flares that occurred in AR 12089\footnote{\url{https://www.solarmonitor.org}}.
For each flare, we check the CME association to the flares by inspecting the {\it SOHO} {\it LASCO} CME catalog\footnote{\url{https://cdaw.gsfc.nasa.gov/CME_list/index.html}}.
In addition, we also visually inspect observations from the Solar Dynamics Observatory~\citep[{\it SDO};][]{Pesnell_2012} to check the signature of CMEs, e.g., coronal dimmings~\citep{Wang_2011}.
We use the UV and EUV observations from the Atmospheric Imaging Assembly~\citep[AIA;][]{Lemen_etal_2012} aboard the {\it SDO} to check the eruption details.
The data have a plate scale of 0\farcs 6 and a cadence of up to 12-s.
Five channels of AIA, 1600, 211, 304, 131, and 94~\AA~are mainly used.
We focus on three events from the AR: two eruptive flares occurred within five hours after the emergence of the AR, along with the largest confined flare.
NOAA categorizes the third flare event as an M1.1 class one.
However, the {\it GOES} SXR data have a gap during this event, therefore we employ the integrated AIA 131~\AA~flux to compare the level of the three flares.
We define the begin, peak, and end times of the flares using the 131~\AA~flux following the criteria used for defining the {\it GOES} X-ray events\footnote{\url{https://www.swpc.noaa.gov/products/goes-x-ray-flux}}: the onset time of an event is defined as the first minute of a consecutive 4-minute period within which the flux exhibits a steep monotonic increase.
The event's maximum time corresponds to the minute when the flux peaks, while the end time is identified when the flux level decreases to a point halfway between the maximum flux and the pre-flare background level.

Based on the vector magnetograms provided by the Helioseismic and Magnetic Imager~\citep[HMI;][]{Scherrer_etal_2012} onboard {\it SDO}, we calculate a few parameters of the AR to investigate the magnetic environment of the AR.
The data segment called Space-Weather HMI Active Region Patches~\citep[SHARP;][]{Bobra_2014} is mainly used. 
We first calculate the total unsigned flux ($\Phi_{\mathrm{AR}}$) of the AR by the formula $\Phi_{\mathrm{AR}} = \sum |B_z| dA$.
To demarcate the core field of the AR, we employ the method proposed by \citet{Sun_2015} to identify a flaring polarity inversion line (FPIL) mask.
Firstly, we choose the polarity inversion line (PIL) pixels from a smoothed map of the vertical field $B_{z}$, then dilate them with a circular kernel which has a radius of 5 pixels (about 1.8 Mm).
Then, we identify flare ribbons in the 1600~\AA~image provided by AIA. 
Flare ribbons are isolated based on an intensity criterion and dilated with a large kernel which has a radius of 15 pixels (about 5.4 Mm).
The region where the dilated PIL intersects with the dilated flare ribbons region is identified as the FPIL. 
Our conclusions are not affected if we use different values of radius.
We then calculate the mean characteristic twist parameter within the FPIL mask region ($\alpha_{\mathrm{FPIL}}$) by the formula $\alpha_{\mathrm{FPIL}} = \frac{\int_S J_z(x, y) B_z(x, y) \,dx \,dy}{\int_S B_z^2(x, y) \,dx \,dy}$.
Furthermore, we calculate the new parameter $\alpha_{\mathrm{FPIL}}/{\Phi_{\mathrm{AR}}}$ proposed by~\citet{Li_etal_2022}.

In addition, we extrapolate the 3D coronal magnetic field using the photospheric magnetograms as the bottom boundary to investigate the coronal environment and magnetic topology of the AR. 
We employ the Green’s function method to reconstruct potential field~\citep[PF;][]{Sakurai_1982, Boocock_2019}, based on which we can calculate the decay index ($n = -\frac{d(\ln B_{P})}{d(\ln h)}$) of the AR to discover the constraint of the background field.
Here, $B_{P}$ represents the external potential field, and $h$ denotes the height from the solar surface. 
Finally, we conduct extrapolations of the nonlinear force-free field~\citep[NLFFF;][]{Wiegelmann_2004} for the chosen events and analyze the magnetic topology. 
Using the 3D coronal nonlinear force-free field and potential field, we calculated the magnetic free energy as $E_f = \frac{1}{8\pi} \int_{V} (B_N^2 - B_P^2) dV$.
Here, $B_{N}$ is the magnetic field from NLFFF extrapolation, and $B_{P}$ denotes the potential field.

\section{RESULTS}\label{sec:result}   

\subsection{Detail of Flares}\label{sec:detail}

\begin{figure*}
\begin{center}
\includegraphics[width=1\hsize]{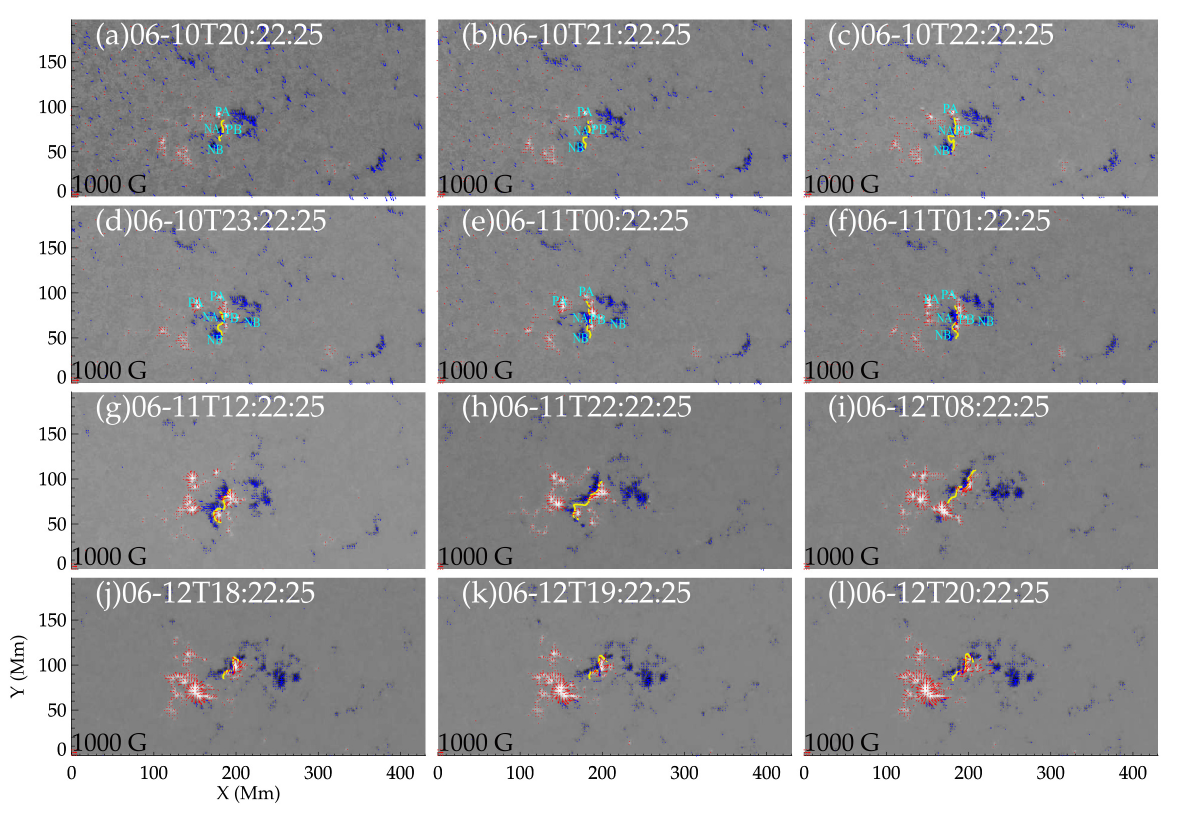} 
\caption{Overview of the evolution of the vector magnetic field of AR 12089. The background is $B_{r}$ component plotted in a dynamic range of ±1000 Gauss, with white (black) regions for the positive (negative) $B_{r}$. 
Red (blue) arrows show the horizontal field component $B_{h}$ that originates from the positive (negative) $B_{r}$ region. 
Yellow lines show the PIL.
An animation lasting from 2014 June 10, 20:22 UT to 2014 June 12, 21:22 UT is available online. 
}\label{fig:mag} 
\end{center}
\end{figure*}

AR 12089 emerged on 2014 June 10 during the Solar Cycle 24.
It produced 11 flares, with only one classified as an M-class flare, while the others were C-class flares. 
Figure~\ref{fig:mag} shows the evolution of the photospheric vector magnetic field of the AR. 
AR 12089 had a multipolar magnetic configuration, consisting of a group of smaller magnetic polarities, PA and NA, as well as a group of slightly larger polarities, PB and NB, in the very early emergence phase (Figure~\ref{fig:mag}(a)-(c)). 
The polarities PA and NB underwent a second phase of emergence (Figure~\ref{fig:mag}(d)-(f)).
Along the PIL (yellow line in Figure~\ref{fig:mag}), the polarities exhibited shearing and converging motions throughout the relevant period (see the movie associated with Figure~\ref{fig:mag}), which were supposed to be associated with flux cancellation~\citep{van Ballegooijen_1989}.
After emergence, the positive polarity PA moved northeastward slightly, while PB  remained stationary.
The negative polarity NA emerged and moved near the positive polarity PB from 20:22 UT on June 10 to 01:22 UT on June 11, after which NA and NB expanded and merged. 
Since 12:22 UT on June 11 (Figure~\ref{fig:mag}(g)), it became difficult to distinguish distinct, same-signed magnetic polarities along the PIL.

\begin{figure*}
\begin{center}
\includegraphics[width=1.0\hsize]{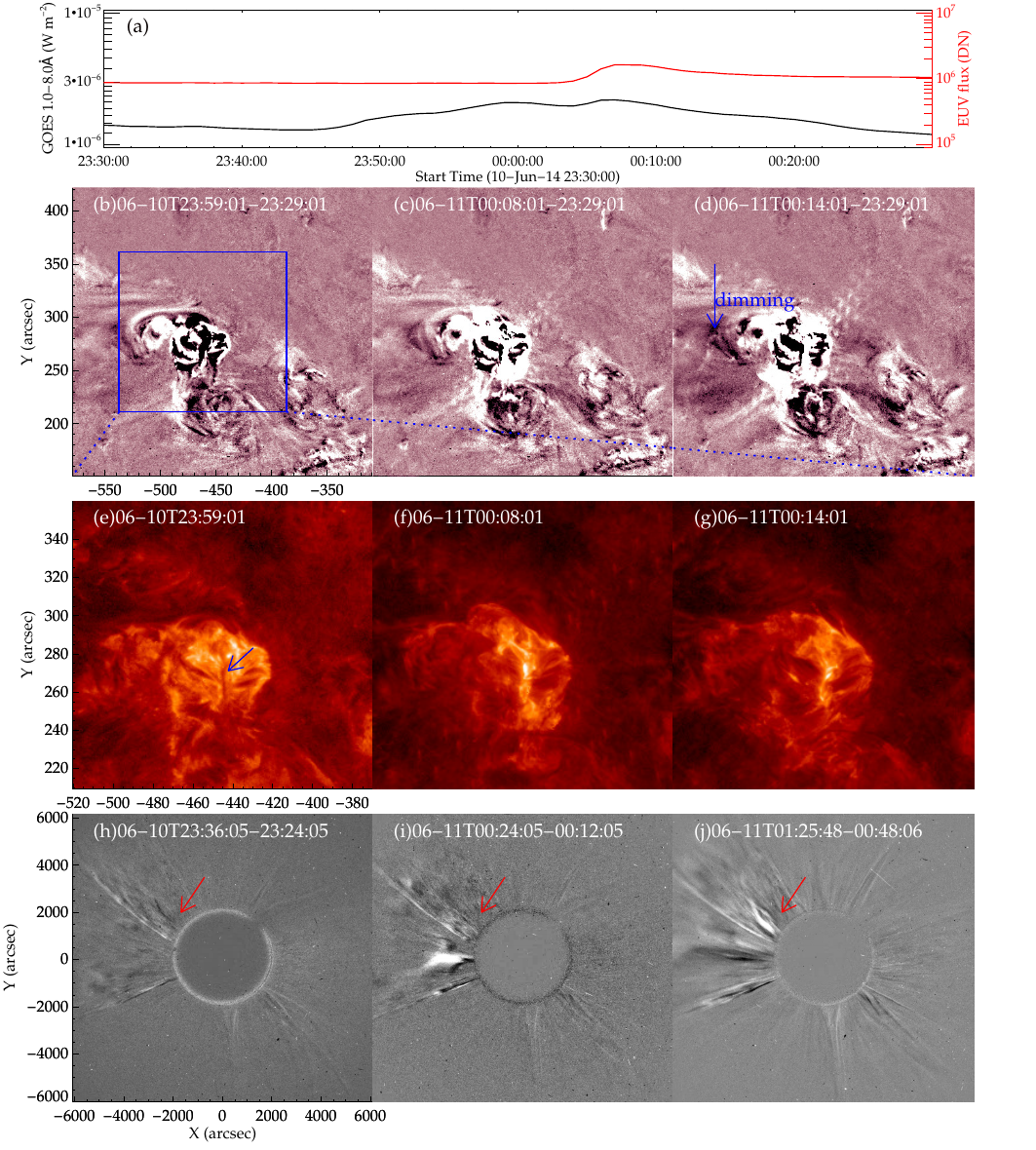} 
\caption{Overview of the C2.1 solar ﬂare in AR 12089, which started at 00:03 UT on 2014 June 11, and ended at 00:14 UT.
(a) {\it GOES} soft X-ray light curves (black line) and the total EUV flux of AIA 131~\AA~(red line). 
(b)-(d) AIA 211~\AA~base-difference images. 
The blue box marks the field of view (FOV) used for panels (e)-(g). 
(e)-(g) AIA 304~\AA~images, showing the filament and eruption.
The blue arrow points to the filament.
(h)-(j) The {\it LASCO} C2 images. The red arrows denote the CME structure.
An animation lasting from 23:30 UT to 00:30 UT is available online. 
}\label{fig:flare1} 
\end{center}
\end{figure*}

For each selected flare, we examine the eruption details.
The first event was a small flare accompanied by a CME.
The flare is well observed by {\it SDO}/AIA. 
{\it GOES} soft X-ray (SXR) curve (Figure~\ref{fig:flare1}(a)) shows that this is a C2.1 class flare.
We also use integrated AIA 131~\AA~flux of the region to check the intensity of the flare. 
The {\it GOES} SXR emission during the flare started to rise at 23:46 UT on 2014 June 10, peaked at 00:08 UT, and lasted until 00:14 UT.  
However, the AIA 131~\AA~flux increase was somewhat later, starting at 00:03 UT on June 11. 
This might be due to the occurrence of another flare (S17E74) on the solar disk during the period from 23:46 UT to 00:03 UT, which is irrelevant to the flare we studied.
Following the definition of {\it GOES} SXR flare, we designate June 11, 00:03 UT as the onset time of the first event.
Figure~\ref{fig:flare1}(b)-~\ref{fig:flare1}(d) display the AIA 211~\AA~base-difference images.  
We can see the obvious dimming (Figure~\ref{fig:flare1}(d)) located on both sides of the flaring region.
As suggested by previous studies~\citep{Thompson_etal_2000}, these dimming regions indicated the occurrence of a CME.
The ejection of emitting plasma during a CME caused darkening in and near the erupting AR, manifested as dimming regions.
Before the flare, we can see a filament exists along the PIL in the central part of the AR clearly in AIA 304~\AA~images (Figure~\ref{fig:flare1}(e)), which is also somewhat visible in AIA 94~\AA~and 171~\AA~images. 
At about 00:04 UT on June 11, the filament started to rise, and then erupted violently to the northeast, culminating in a partial ejection. 
Meanwhile, beneath the rising filament, the intense EUV brightening can be seen (Figure~\ref{fig:flare1}(f), see also in the associated movie). 
Soon after the disk eruption, a CME was observed in {\it LASCO}/C2 white-light images (pointed out by the red arrow in Figure~\ref{fig:flare1}(h)-(j), see also in the associated movie).

\begin{figure*}
\begin{center}
\includegraphics[width=1.0\hsize]{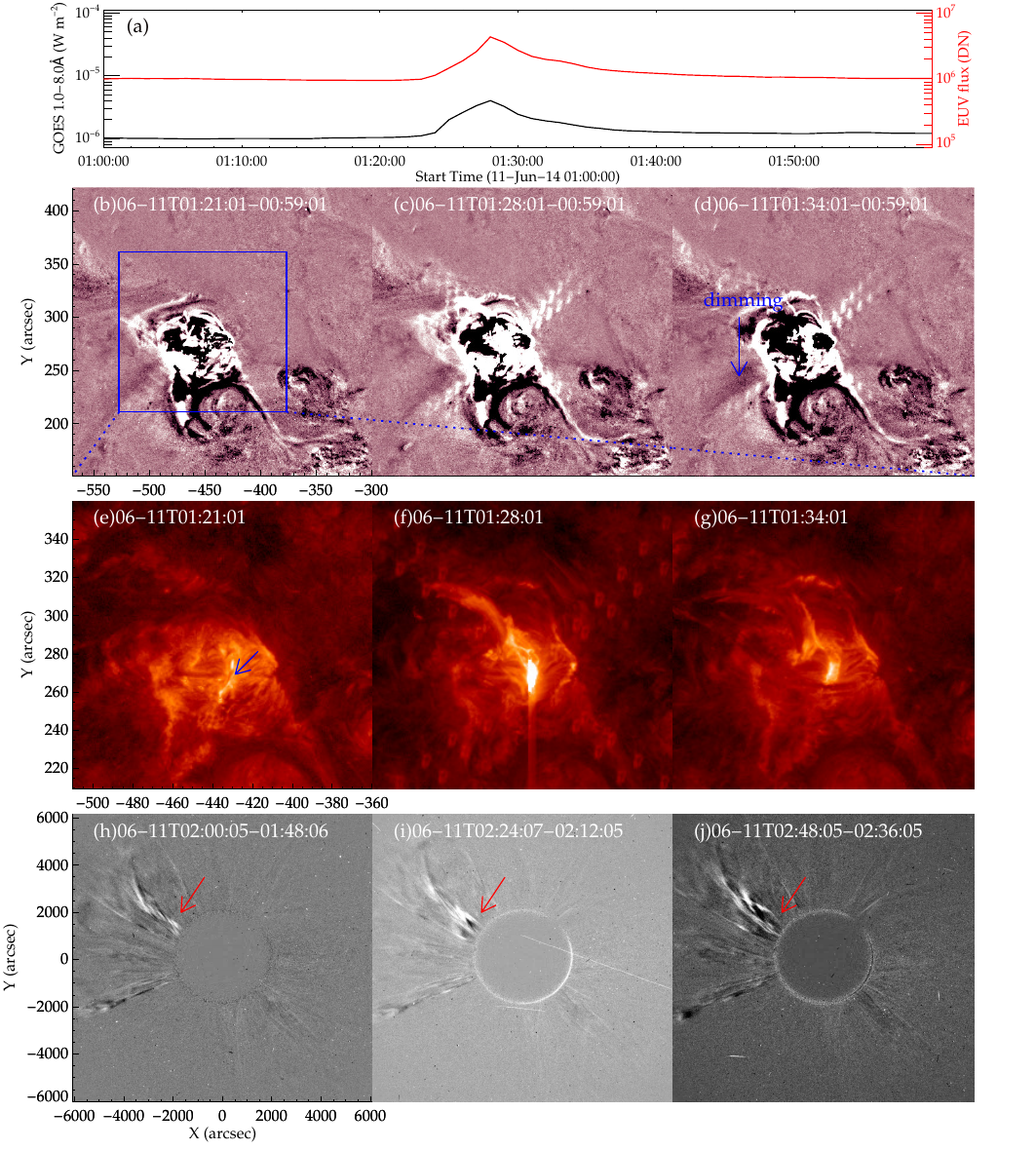} 
\caption{Overview of the C3.9 ﬂare in AR 12089, which started at 01:22 UT on 2014 June 11, and ended at 01:30 UT.
Similar layout as Figure 2.  
An animation lasting from 01:00 UT to 02:00 UT is available online.}\label{fig:flare2} 
\end{center}
\end{figure*}

About an hour after the first flare event, the second C3.9 class flare began at 01:22 UT on June 11. 
It was accompanied by a CME as well.
We use {\it GOES} SXR, AIA 211, 304~\AA~, and {\it LASCO}/C2 images to show the flare progress (Figure~\ref{fig:flare2}).
{\it GOES} SXR curve shows this event began at 01:22 UT, peaked at 01:28 UT, and ended at 01:30 UT, while EUV flux has a similar curve (Figure~\ref{fig:flare2}(a)).
The coronal dimming appeared at both sides of the flaring region in AIA 211~\AA~images (see also in Figure~\ref{fig:flare2} associated movie), which suggested that there was a CME.
The filament, which partially remained after the first eruption, was reconstructed to a new filament in its original location before the second flare (pointed out by the blue arrow in Figure~\ref{fig:flare2}(e)).
At 01:24 UT, the reconstructed filament erupted from the PIL region as shown in AIA 304~\AA~images (Figure~\ref{fig:flare2}(f)-(g)).
A CME that was very likely associated with this eruption was visible in {\it LASCO}/C2 images (pointed out by the red arrow in Figure~\ref{fig:flare2}(h)-(j)).

\begin{figure*}
\begin{center}
\includegraphics[width=1\hsize]{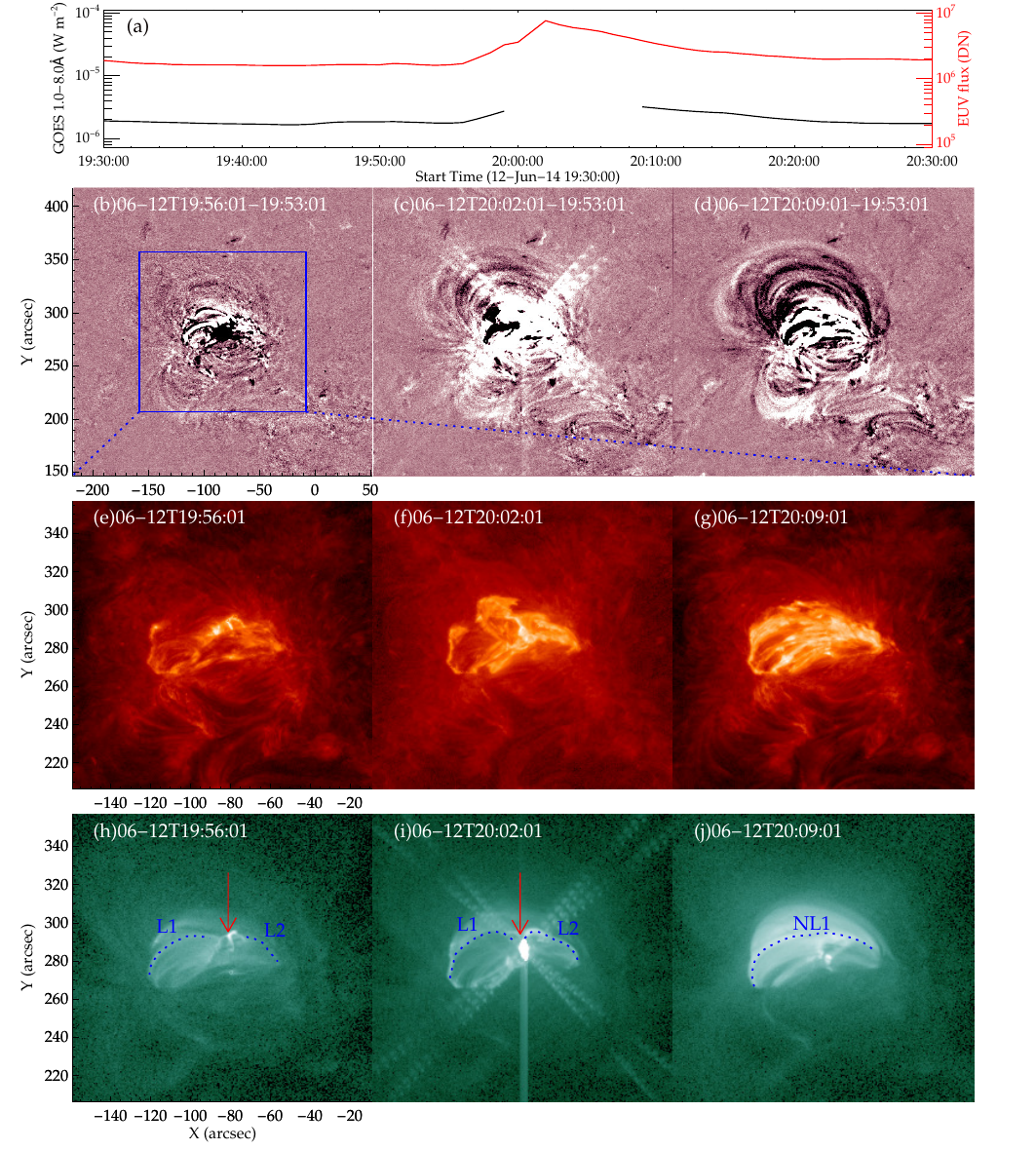} 
\caption{Overview of the M1.1 solar ﬂare in AR 12089, which started at 19:56 UT on 2014 June 12, and ended at 20:05 UT.
(a) {\it GOES} soft X-ray light curves (black line) and total EUV flux of AIA 131~\AA~(red line). 
(b)-(d) AIA 211~\AA~base-difference image. The black box marks the FOV used for panel (e)-(j). 
(e)-(g) AIA 304~\AA~images, showing the eruption. 
(h)-(j) AIA 94~\AA~images. L1 and L2 in panels (h) and (i) are brightened loop bundles identified in 94~\AA~images prior to the flare.
The red arrows point to the brightenings.
NL1 in panel (j) denotes loop bundles newly formed during the flare.
An animation lasting from 19:30 UT to 20:30 UT is available online. 
}\label{fig:flare3} 
\end{center}
\end{figure*}

The last flare began from 19:56 UT on June 12.
Due to the data gap of {\it GOES} SXR data from 20:00 UT to 20:08 UT, we use the definition of the begin, maximum, and end-time of a {\it GOES} X-ray flare to define the timings of this event from the AIA flux.
The flare started at 19:56 UT, peaked at 20:03 UT, and ended at 20:05 UT (Figure~\ref{fig:flare3}(a)).
The peak AIA  flux of this event was much higher than the flux of the two previous events, which means this flare was more energetic than the previous ones.
As seen from the 211~\AA~base-difference images in Figure~\ref{fig:flare3}(b)-(d) and accompanying animation, there was no significant dimming occurred.
In 304~\AA~images~(Figure~\ref{fig:flare3}(e)-(g)), there was no clear filament observed.
From around 20:00 UT, a gentle eruption occurred.
Soon afterward, at 20:03 UT, there was a slight mass lift, different from the previous events in which the dramatic filament eruptions and much higher mass ejection occurred.
No related CME was found in the {\it SOHO}/{\it LASCO} observations. 
It seems that the flare was not accompanied by a CME.
Therefore, we utilize 94~\AA~images to illustrate this eruption. 
At 19:56 UT, we identify two sets of bright loop bundles labeled as L1 and L2 (Figure~\ref{fig:flare3}(h)).
L1 and L2 were located close to each other and seemed to reconnect at about 19:56 UT (red arrow in Figure~\ref{fig:flare3}(h)).
Another brightening appeared at 20:01 UT (red arrow in Figure~\ref{fig:flare3}(i)).
Later on, new loop bundles labeled NL1 appeared at 20:09 UT (Figure~\ref{fig:flare3}(j)), implying the reconfiguration of magnetic fields resulted from magnetic reconnection.
The brightenings at the eastern footpoints of loop bundles L1 and western footpoints of L2 exhibited apparent slipping motions in 94 A images (see Figure~\ref{fig:flare3} associated movie). 
This process is consistent with slipping reconnection occurring at the quasi-separatrix layers~\citep[QSLs,][]{Demoulin_etal_1996}, in which magnetic field lines passing through QSLs successively reconnect with neighboring field lines~\citep{Priest_1995, Aulanier_etal_2006, Dudik_etal_2014}.

\begin{figure*}
\begin{center}
\includegraphics[width=1.0\hsize]{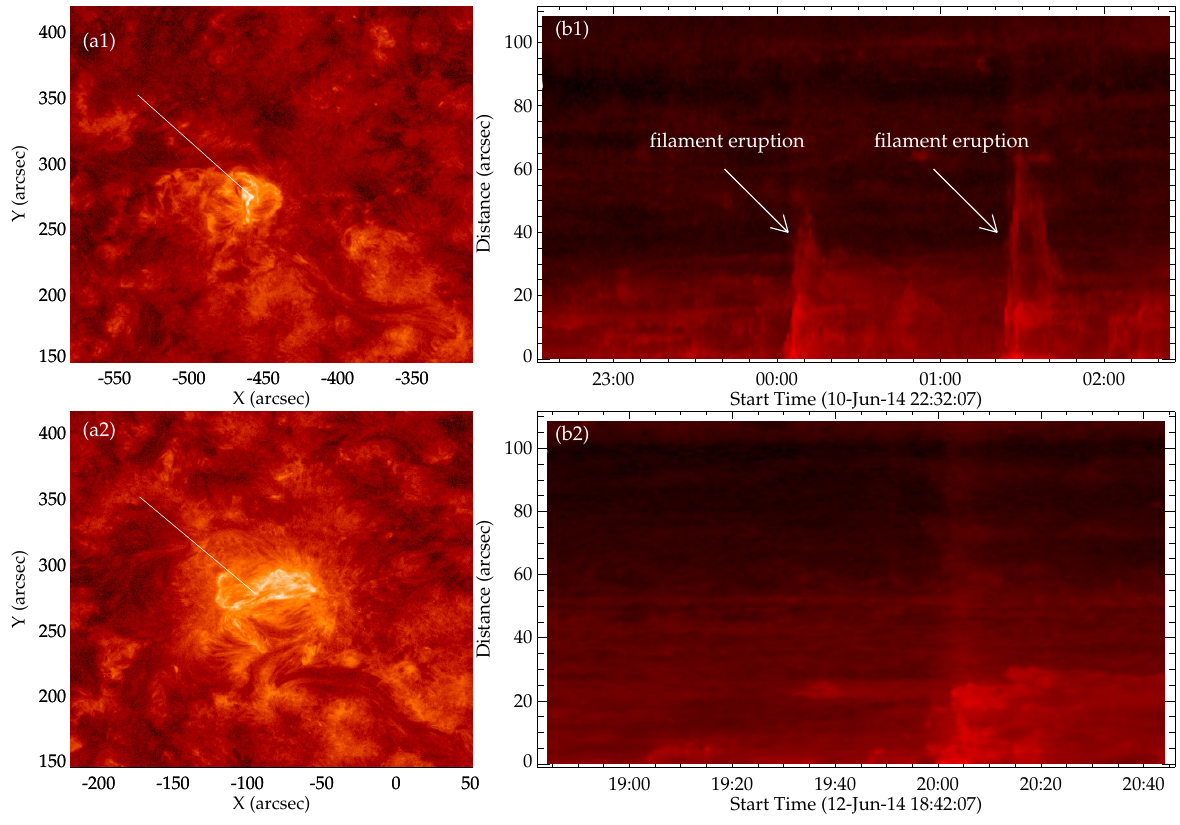} 
\caption{Time distance maps of AIA 304 Å images for the slits shown in panels (a1) and (a2). 
The white arrows in panel (b1) point to the filament eruption.
}\label{fig:jmap} 
\end{center}
\end{figure*}

We draw a slice (white line in Figure~\ref{fig:jmap}(a1), (a2)) perpendicular to the PIL and plot the time-distance diagram of the slice in 304~\AA~images to further see the eruption details of the flares (Figure~\ref{fig:jmap}).  
From the time-distance plots, we note that the material ejection process of the first two eruptions could be followed.
During the first flare, since about 00:03 UT, the filament started to rise and then erupted at about 00:08 UT. 
During the second flare, in the same location, at about 01:21 UT, the reconstructed filament rose again and erupted at about 01:28 UT (Figure~\ref{fig:jmap}(b1)).
On the contrary, the confined flare did not produce a significant material ejection (Figure~\ref{fig:jmap}(b2)). 
Only brightening appeared from about 20:00 UT to 20:20 UT.
This result also confirms that the first two flares were eruptive, while the third flare was confined.
The AR produced two successive CMEs within five hours after its emergence, which is rarely observed.

\subsection{Magnetic Parameters}\label{sec:mag}

\begin{figure*}
\begin{center}
\epsscale{1.18}
\plotone{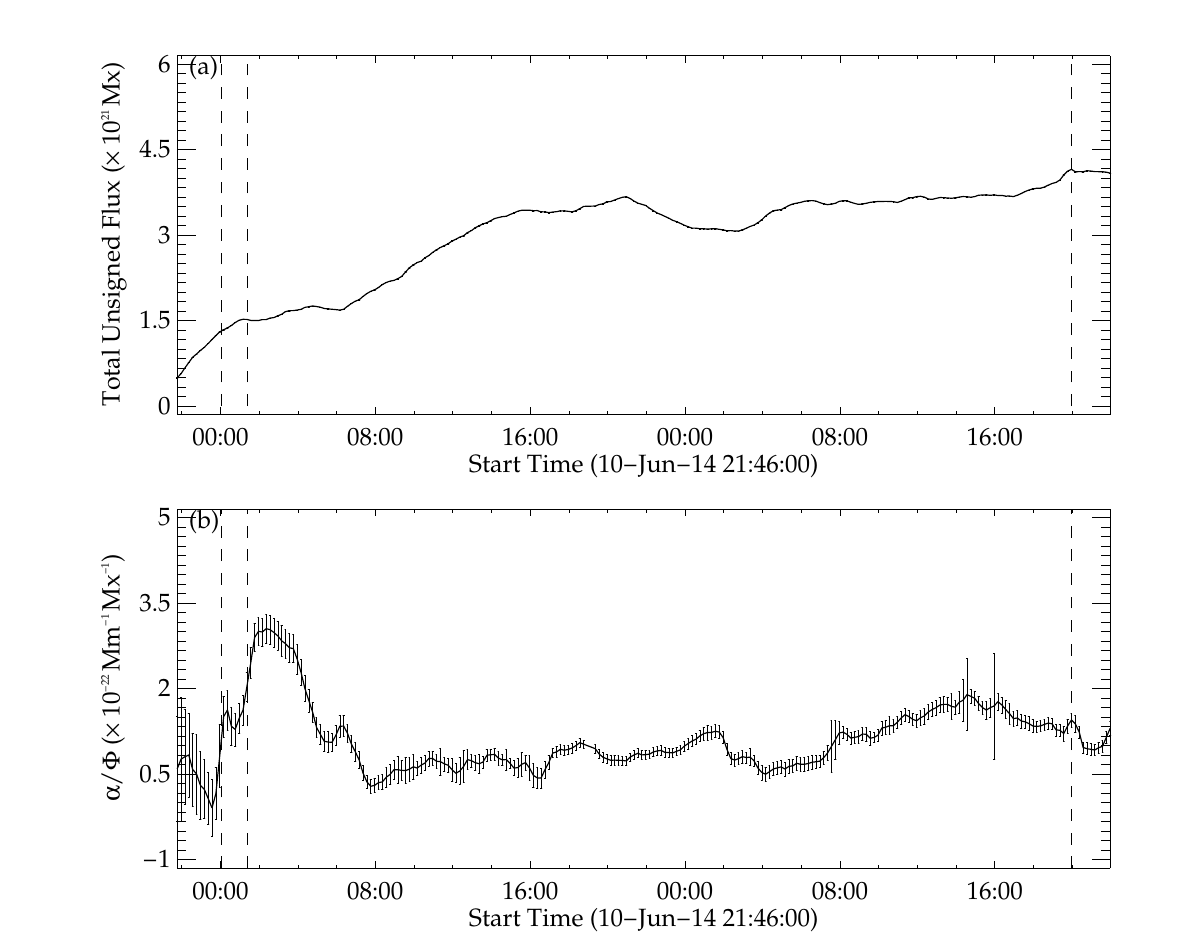}
\caption{(a) Temporal evolution of total unsigned magnetic flux (black line) and magnetic free energy (red scatter points).
(b) Temporal evolution of  $\alpha_{\mathrm{FPIL}} \slash \Phi_{\mathrm{AR}}$.
The three vertical dashed lines indicate the onset time of the flares.
In each panel, a three-point smoothing operation is performed on the data for better visibility of underlying trends.
}\label{fig:envir} 
\end{center}
\end{figure*}

To investigate why the AR produced eruptive flares so early, we calculate a few magnetic parameters that mainly quantify the magnetic environment of the AR, such as the non-potentiality and background field constraint, using the photospheric vector magnetograms and the extrapolated fields.
The difference in the magnetic environment may be crucial in determining whether a flare is associated with a CME.
We calculate the total unsigned magnetic flux ($\Phi_{\mathrm{AR}}$), magnetic free energy ($E_f$), and the mean characteristic twist parameter within the FPIL mask region ($\alpha_{\mathrm{FPIL}}$) versus $\Phi_{\mathrm{AR}}$ (Figure~\ref{fig:envir}).
The error propagated from the error segment of the photospheric magnetic field data~\citep{Bobra_2014} are overplotted as error bars, and they are relatively small compared to the $\Phi_{\mathrm{AR}}$ values.
For the $E_f$, we calculate it in the period that includes both the pre- and post-flare phases for the three flares, as well as at several time points between June 11 and 12, to ensure the continuity of the data and demonstrate the evolutionary trend.

Figure~\ref{fig:envir}(a) shows the evolution of $\Phi_{\mathrm{AR}}$ and $E_f$, spanning from the emergence of the AR to the end of the third flare.
Throughout the process, the magnetic flux increases from $4.96 \times 10^{20}$ Mx to about $4.18 \times 10^{21}$ Mx.
It was on the order of $10^{21}$ Mx when producing the first two flares, which is at least an order of magnitude smaller than the value of the magnetic flux of the commonly seen CME-producing ARs shown in the statistical research~\citep{Chen_etal_2011}.
Similarly, the value of $E_f$ shows an overall increasing trend.
During the eruptive flares, the values of $E_f$ are on the order of $10^{30}$ erg, which is also an order of magnitude smaller than the minimum free energy of the CME-producing ARs in the statistical study~\citep{Vasantharaju_etal_2018}.

On the other hand, the $\alpha_{\mathrm{FPIL}}/{\Phi_{\mathrm{AR}}}$ shows different tendency (Figure~\ref{fig:envir}(b)).
It can distinguish whether a flare is erupted or confined in our study. 
The value of $\alpha_{\mathrm{FPIL}}/{\Phi_{\mathrm{AR}}}$ increases fast since the emergence of the AR. 
We calculate the average value of the two nearest times before and after the flare onset as a reference for the value of $\alpha_{\mathrm{FPIL}}/{\Phi_{\mathrm{AR}}}$ at the beginning of the flare.
At around 00:03 UT on June 11, the first flare onset, the $\alpha_{\mathrm{FPIL}}/{\Phi_{\mathrm{AR}}}$ is $1.43 \times 10^{-22} \, \mathrm{Mm}^{-1} \, \mathrm{Mx}^{-1}$. 
When the second flare commenced at 01:22 UT, the $\alpha_{\mathrm{FPIL}}/{\Phi_{\mathrm{AR}}}$ has increased to $2.15 \times 10^{-22} \, \mathrm{Mm}^{-1} \, \mathrm{Mx}^{-1}$. 
Later, the value of $\alpha_{\mathrm{FPIL}}/{\Phi_{\mathrm{AR}}}$ shows a general trend of decrease as the flux emergence goes on.
The M1.1 confined flare started at 19:56 UT on June 12, with an $\alpha_{\mathrm{FPIL}}/{\Phi_{\mathrm{AR}}}$ value of $1.34 \times 10^{-22} \, \mathrm{Mm}^{-1} \, \mathrm{Mx}^{-1}$. 
In general, the value of $\alpha_{\mathrm{FPIL}}/{\Phi_{\mathrm{AR}}}$ rapidly increases after the emergence of the AR, reaching its peak shortly after the end of the second flare. 
The value during the two eruptive flares is larger than that in the confined flare.
The parameter considers the magnetic non-potentiality of the core region and the constraining effect of the background field at the same time.
A higher value of this parameter indicates a greater possibility of eruptive flares.
This is consistent with the observation that the first two flares are eruptive, while the third is confined.
In previous research,~\citet{Li_etal_2022} examined 106 flares of {\it GOES} class larger than M1.0 and found that the critical value of $\alpha_{\mathrm{FPIL}}/{\Phi_{\mathrm{AR}}}$ to distinguish most eruptive flares from the confined flares is $2.2 \times 10^{-24} \, \mathrm{Mm}^{-1} \, \mathrm{Mx}^{-1}$. 
In our case, the value for eruptive flares is two orders of magnitude larger than the critical value in their study.  
Moreover, the maximum value in their statistical study is on the order of $10^{-23} \, \mathrm{Mm}^{-1} \, \mathrm{Mx}^{-1}$, one order of magnitude smaller than our results.
The $\alpha_{\mathrm{FPIL}}/{\Phi_{\mathrm{AR}}}$ values of AR 12089 here deviate from the statistical values, which might be attributed to the small size of the AR. 
The $\Phi_{\mathrm{AR}}$ in AR 12089 is at least an order of magnitude smaller than that in the statistical study.
As a result, the $\alpha_{\mathrm{FPIL}}/{\Phi_{\mathrm{AR}}}$ is at least one order of magnitude larger than that in~\citet{Li_etal_2022}.

\begin{figure*}
\begin{center}
\epsscale{1.1}
\plotone{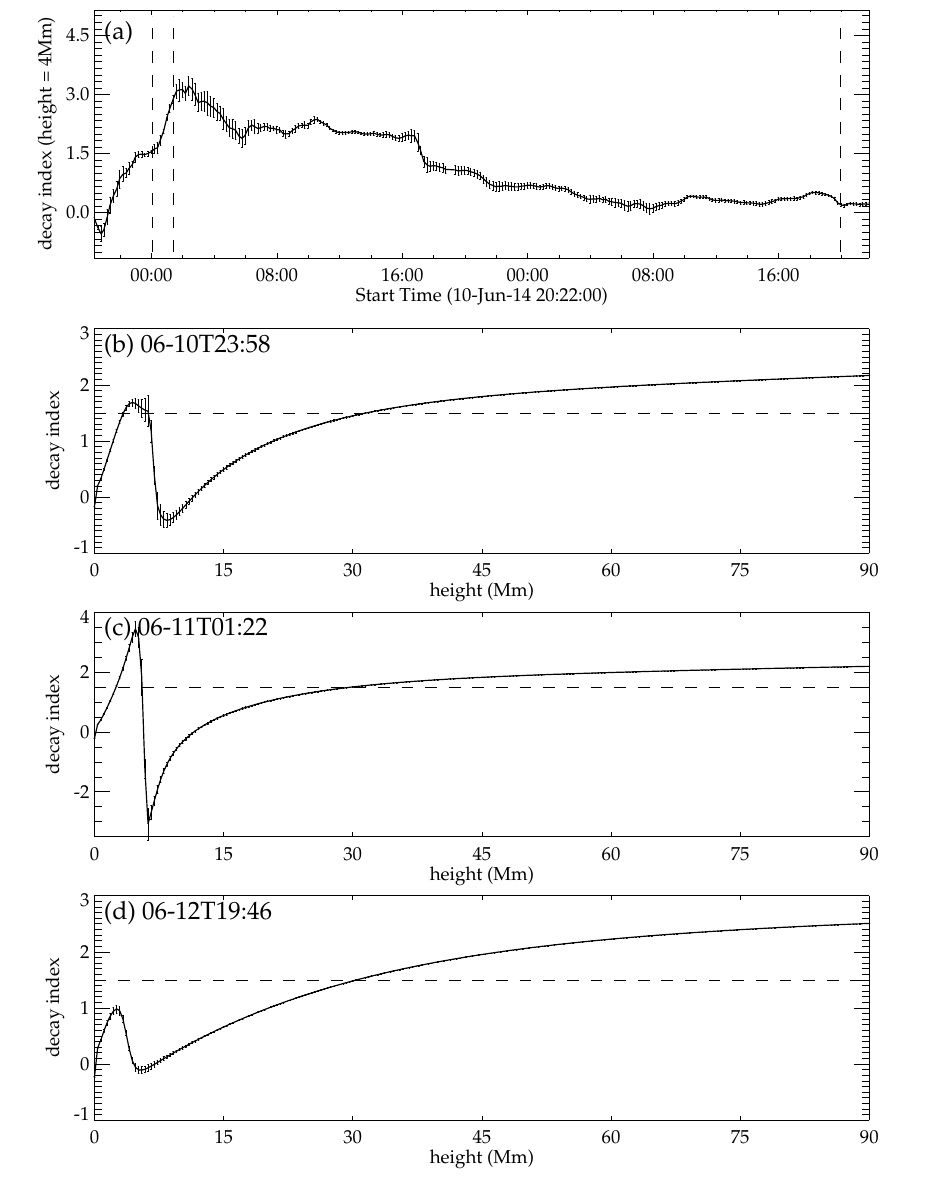}
\caption{(a) Temporal evolution of decay index averaged at 4 Mm above the PIL. 
A three-point smoothing operation is performed on the data for better visibility of underlying trends. 
The vertical dashed lines indicate the flares' onset. 
(b)-(d) Distribution of the mean decay index above the PIL before the onset of flares. 
The horizontal dashed lines mark the critical value of 1.5.
}\label{fig:decay_index} 
\end{center}
\end{figure*}

Furthermore, we compute the decay index $n$ to discover the constraint above the PIL. 
In torus instability theory, the critical value of the decay index for an eruption to occur is 1.5.
This means that when the decay index exceeds the threshold, the constraint in the corona decreases rapidly enough, that small perturbations can lead to the CME seed such as a flux rope erupting out more easily~\citep{Kliem_2006}.
Figure~\ref{fig:decay_index}(a) shows the temporal evolution of the decay index averaged at 4 Mm above the photospheric PIL.  
The decay index before the onset of the three flares is 1.66, 2.80, and 0.28, respectively.
Figure~\ref{fig:decay_index}(b)-(d) displays the distribution of the decay index with height prior to the flares.
At each height, the decay index is averaged along the PIL.
The standard error is overplotted as error bars.
In all three events, the $n$-profiles exhibit a saddle-like shape, with a local valley of $n$ value enclosed by two regions of higher $n$ value, which usually appears in the quadrupolar magnetic configuration~\citep{Filippov_2020, Luo_2022}.
Before the two eruptive flares, in the range of 0-15 Mm, i.e., the low corona, we can see the decay index reached 1.5 at around 4.5 Mm.
At the same time, the local maximum of $n$ is 1.69 and 3.45, which suggests that the flux rope (if there is any) would be subject to torus instability when rising to this height.
However, before the confined flare, the local maximum decay index is only 0.98, less than 1.5. 
This suggests that during the third event, the constraint was large so that the rising flux rope (if there is any) was hard to evolve into a CME. 
It is worth noting that the bottom of the saddle-like profile of $n$ exhibits negative values. 
This might be attributed to the possible presence of magnetic null-point in the core region of the AR, similar to what~\citet{Mitra_etal_2022} reported in a quadrupolar magnetic configuration.

Here we only calculate the decay index vertically above the PIL and study its distribution before the onset of flares. 
It should be noted that the constraining force above the flux rope may change dynamically with the movement of the flux rope during the eruption, so that the static decay index before the flare may not be adequate to determine the whole eruption process.
A laboratory experiment revealed a failed torus regime, which found that the change in the toroidal field tension force induced by the interaction between the external toroidal field (the guide field) and the flux rope poloidal current could also restrict the eruption~\citep{Myers_etal_2015}.
The laboratory failed torus event is found associated with the conversion of the poloidal field into the toroidal field when the magnetic flux rope erupts~\citep{Myers_etal_2015}. 
The conversion may decrease the twist number of the flux rope, manifested as the untwisting rotation of the flux rope. 
Indeed, in solar observations, the failed torus event is often associated with large-angle rotations~\citep{Zhou_etal_2019, Zhang_etal_2024}.
In addition to the observations, MHD simulations also successfully reproduced the failed torus process,  revealing the underlying physical details~\citep{Jiang_etal_2023, Guo_etal_2024}.
Different from the reported failed torus events, in our eruptive flares, the filament~\citep[seen as the proxy of the flux rope,][]{Cheng_etal_2017} erupted directly without any clear rotation (see Figure~\ref{fig:flare1},~\ref{fig:flare2} and the related movies).
In the confined flare, no corresponding filament or hot channel~\citep[both seen as the proxy of the flux rope,][]{Cheng_etal_2017} was observed. 
Only simple slipping reconnection occurred (see Figure~\ref{fig:flare3} and the related movie). 
It is more likely that the failed torus scenario does not occur in our case.

\subsection{Evolution of the Flux Rope}\label{sec:flux ropes}

\begin{figure*}
\begin{center}
\epsscale{1.1}
\plotone{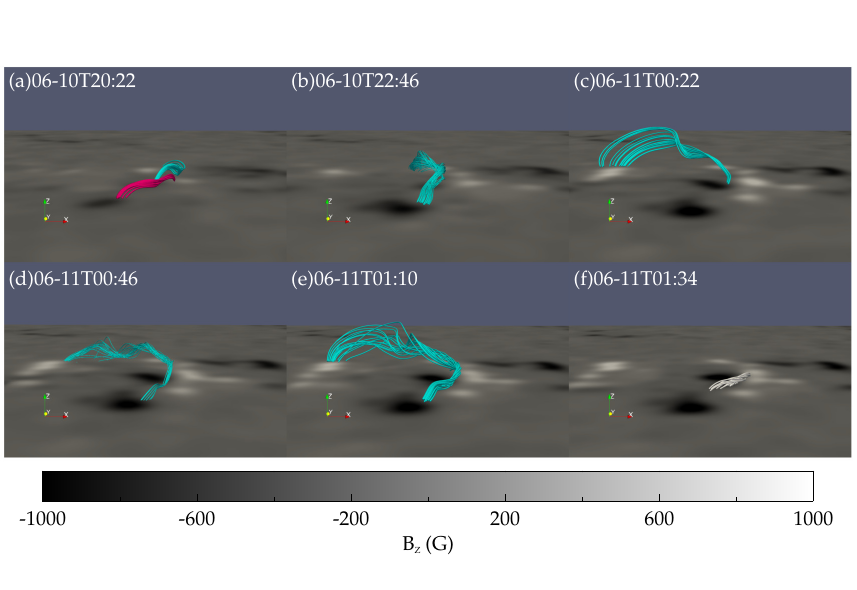}
\caption{The evolution of coronal structure above the core region of the AR. 
The time for panels (a)-(b) are before the onset of the first flare, (c)-(e) are between the end of the first flare and the onset of the second flare, and (f) is after the end of the second flare.
The sheared flux arcades and flux ropes are marked in magenta and cyan.
The grey lines in panel (f) represent the remaining magnetic loops after the second flare.
}\label{fig:nlf} 
\end{center}
\end{figure*}

As presented in Section~\ref{sec:detail}, the former two flares are associated with filament eruptions.
To investigate whether the filaments are associated with magnetic flux ropes or simply highly sheared structures, we use the photospheric vector magnetograms to extrapolate the coronal NLFFF~\citep{Wiegelmann_2004}.
The NLFFF extrapolation could reproduce magnetic flux ropes and other magnetic topological characteristics of the AR.
We mainly focus on the core coronal structure above the PIL. 
Figure~\ref{fig:nlf} shows the evolution of the core structure from the emergence of the AR (20:22 UT on June 10) to the end of the second eruptive flare (01:34 UT on June 11).
At 20:22 UT, no flux rope was found. 
The core structure consisted of two groups of sheared magnetic arcades across the PIL, connecting magnetic polarities NA and PA (marked in cyan) and NB and PB (marked in magenta), respectively (Figure~\ref{fig:nlf}(a)). 
Footpoints of the two sheared magnetic arcades, NA and PB, were located close to each other. 
Due to shear and converging motions along the PIL (as seen in Figure~\ref{fig:mag}), the sheared arcades may reconnect gradually to form a magnetic flux rope.
A flux rope connecting PA and NB did appear around 22:46 UT (Figure~\ref{fig:nlf}(b)).
As the mass accumulated along the flux rope, a filament became visible as shown in the AIA 304~\AA~image (Figure~\ref{fig:flare1}(e)).
After the first flare, the flux rope erupted partially, with part of it remaining (Figure~\ref{fig:nlf}(c)). 
Note that the northern footpoint of the remaining part was connected to the far side of PA, suggesting a reconfiguration of the magnetic field accompanied by the flux rope eruption.
The flux rope reformed gradually after the first flare (Figure~\ref{fig:nlf}(d)-(e)).
At 01:22 UT, the second flare occurred, during which the flux rope ejected completely and left only nearly-potential magnetic loops (Figure~\ref{fig:nlf}(f)). 
The results indicate that during the stage when the first two flares occurred, a flux rope was formed repeatedly above the PIL, possibly through flux cancellation along the PIL.
The flux rope ejected out repeatedly and produced successive CMEs.

After the second flare, no flux rope could be found above the PIL, even before the confined flare event on June 12. 
No coherent highly sheared magnetic structures were found either.
Therefore, we do not present the results here.
Combining the EUV observations (Figure~\ref{fig:flare3}), it is more likely that there was no flux rope existing prior to, or formed during the confined flare.
The flare may be a result of simple reconnection between nearby coronal loops, rather than the ejection of a flux rope.

\section{SUMMARY and DISCUSSION}\label{sec:sum}

In this work, we analyze and interpret a rare case involving two C-class eruptive flares and an M-class confined flare that occurred in a newly emerged AR 12089. 
There have been comparative studies of eruptive and confined flares produced in a single AR~\citep[eg.,][]{Cheng_etal_2011, Shen_etal_2011}, but AR 12089 here has a certain distinctiveness. 
We consider the AR as a rare case for two reasons. 
First, it produced eruptive flares within five hours after its emergence, significantly shorter than the typical interval of around 1 day from the AR emergence to the first major eruption as presented in the statistical research~\citep{Liu_etal_2021}. 
Second, the AR was extremely small when producing the eruptive flares. 
It had the magnetic flux on the order of $10^{21}$ Mx and magnetic free energy on the order of $10^{30}$ erg, both of which also fall outside the values in the statistical results~\citep{Chen_etal_2011, Vasantharaju_etal_2018}. 
The analyses of the AR are summarized in the following.
The first eruptive C2.1 class flare occurred less than 4 hours after the AR emerged.
During this event, a pre-existing filament was partially expelled, resulting in a CME.
Within tens of minutes after the first flare, the partially remaining filament was reformed.
About 1.3 hours later, the second eruptive C3.9 class flare occurred, during which the filament erupted out fully, giving rise to a CME.
As a comparative study, a confined M1.1 class flare from the AR is investigated.
No obvious filament structure is found involved in this event.
slipping reconnections between two sets of coronal loop bundles, similar to the type I confined flares proposed by~\citet{Li_etal_2019}, during which no filament participates.

The above eruption process is confirmed by a series of NLFFF extrapolations of the coronal magnetic field. 
The filament is usually seen as a proxy of the flux rope~\citep{Cheng_etal_2017}.
We do find a flux rope in the extrapolated NLFFF before the onset of the first flare, which was possibly formed through flux cancellation along the PIL.
The flux rope partially ejected out during the first flare, leading to a CME.
After the first flare, the remaining sheared structures gradually reformed into a new flux rope.
During the second flare, the flux rope ejected out completely, leaving only nearly-potential magnetic loops.
After the second flare, no flux rope was found above the PIL anymore.
Neither were coherent sheared structures found.
This confirms that no flux rope was involved during the confined flare.

Moreover, we calculate parameters quantifying the magnetic constraint (decay index and total unsigned flux), the non-potentiality (magnetic free energy), and the competition between the core-field non-potentiality and the coronal constraint ($\alpha_{\mathrm{FPIL}}/{\Phi_{\mathrm{AR}}}$) for both the eruptive and confined flares.
For the decay index, we find its profile is saddle-like before the occurrence of the three flare events, with the decay index at the saddle top being 1.69, 3.45, and 0.98, respectively.
The decay index exceeding the critical value of 1.5 in the first two events suggests that the flux rope found in the extrapolated NLFFF would easily evolve into a CME under slight perturbation due to the torus instability mechanism.
For the third event, the decay index below 1.5 suggests a stronger background constraint, implying that even if a flux rope exists, its eruption is not easy to occur. 
Note that previous studies indicate that not all flux ropes can eject out successfully even the decay index exceeds the critical value, which is known as the failed torus scenario~\citep{Myers_etal_2015, Zhou_etal_2019, Zhang_etal_2024}. 
Such failed torus events are found associated with the conversion from poloidal field to toroidal field~\citep{Myers_etal_2015}.
In addition, a non-axisymmetrical force induced by the radial magnetic field can also restrict the eruption, which is another possible reason for failed torus events~\citep{Zhong_etal_2021, Zhang_etal_2024}.
However, observations in AR 12089 did not reveal signatures relevant to the failed torus, such as the rotation of the flux rope during the eruptive flares. 
In the confined flare, no flux rope was found to participate. 
It is more likely that in our case, the failed torus scenario does not occur.
The decay index distribution before the onset of flares can determine the eruption characteristics to a large extent here.

Another interesting finding about the decay index in our case is that for all three events, there are negative values of the decay index at the saddle bottom.
Negative decay index is rarely reported.
Only one research suggests it is associated with the coronal null-point existing above the torus-unstable region in a multi-polar region~\citep{Mitra_etal_2022}.
Although the negative decay index seems to play no role in the eruptive characteristics of the flares here, its formation deserves further analysis in a larger sample.

For the $\Phi_{\mathrm{AR}}$ and the $E_f$, they both show an overall increasing trend.
Notably, their values during the eruptive flares are on the order of $10^{21}$ Mx and $10^{30}$ erg, respectively, both being one order of magnitude lower than the common values reported in statistical studies~\citep{Chen_etal_2011, Vasantharaju_etal_2018}.
The value of $\alpha_{\mathrm{FPIL}}/{\Phi_{\mathrm{AR}}}$ exhibits a rapid increase after the emergence of the AR, reaching its peak shortly after the end of the second flare.
The value is higher during eruptive flares compared to that in the confined flare, suggesting that the relative measure of the AR's non-potentiality over the background magnetic field constraint is greater during eruptive flares.
Moreover, in our study, the magnitude of $\alpha_{\mathrm{FPIL}}/{\Phi_{\mathrm{AR}}}$ is at least one order higher than the value of $\alpha_{\mathrm{FPIL}}/{\Phi_{\mathrm{AR}}}$ for the eruptive flares in the statistical study~\citep{Li_etal_2022}.
This is mainly attributed to the fact that the extremely small AR has a much lower $\Phi_{\mathrm{AR}}$ than that in the statistical study, resulting in a much higher  $\alpha_{\mathrm{FPIL}}/{\Phi_{\mathrm{AR}}}$.
These results complement the sample and extend our understanding of the relative measure $\alpha_{\mathrm{FPIL}}/{\Phi_{\mathrm{AR}}}$, confirming that compared to the extensive-type parameters such as $\Phi_{\mathrm{AR}}$ and $E_f$, the $\alpha_{\mathrm{FPIL}}/{\Phi_{\mathrm{AR}}}$, as a relative measure, is more effective in reflecting the ability of the AR to produce eruptive flares.

In summary, we focused on a rare active region, AR 12089, in this study.
The AR produced eruptive flares within only five hours after its emergence, which is rarely observed. 
Moreover, the magnetic parameters of this AR, which measure the AR size, overall non-potentiality, and the ratio between non-potentiality and overall constraint, all fall outside of the typical statistical ranges~\citep{Chen_etal_2011, Vasantharaju_etal_2018, Li_etal_2022}.
The former two parameters are exceptionally small, while the latter is notably large.
Combining the EUV observations and the magnetic properties of the AR, we conclude that for this exceptional AR, the eruptive flares were the results of the eruption of a magnetic flux rope in the core region and a weak constraint above the core region, while the confined flare was a result of simple reconnection between nearby coronal loops.
This case demonstrates that even for an extremely small, new-born AR, it has the potential to generate eruptive flares if the magnetic non-potentiality within the AR core region dominates over the background field constraint above the core region.

\clearpage

\clearpage

\clearpage

\clearpage

\clearpage

\clearpage

\clearpage

\clearpage

\clearpage

\begin{acknowledgments}
We acknowledge the {\it SDO}, {\it SOHO}, and {\it GOES} missions for providing quality observations.
Lijuan Liu acknowledges the support received from the National Natural Science Foundation of China (NSFC grant no. 12273123) and from the Guangdong Basic and Applied Basic Research Foundation (2022A1515011548, 2023A1515030185).
\end{acknowledgments}


\bibliographystyle{aasjournal}

\end{document}